\documentclass{aa}
\usepackage{graphicx}
\usepackage{savesym}
\usepackage{amsmath}
\usepackage{txfonts}
\usepackage{color}
\usepackage{paralist}
\usepackage{wasysym}
\usepackage[normalem]{ulem}

\usepackage{listings}
\usepackage{natbib,twoopt}
\usepackage[breaklinks=true]{hyperref} 
\bibpunct{(}{)}{;}{a}{}{,}             
\makeatletter
  \newcommandtwoopt{\citeads}[3][][]{\href{http://adsabs.harvard.edu/abs/#3}%
    {\def\hyper@linkstart##1##2{}%
     \let\hyper@linkend\@empty\citealp[#1][#2]{#3}}}
  \newcommandtwoopt{\citepads}[3][][]{\href{http://adsabs.harvard.edu/abs/#3}%
    {\def\hyper@linkstart##1##2{}%
     \let\hyper@linkend\@empty\citep[#1][#2]{#3}}}
  \newcommandtwoopt{\citetads}[3][][]{\href{http://adsabs.harvard.edu/abs/#3}%
    {\def\hyper@linkstart##1##2{}%
     \let\hyper@linkend\@empty\citet[#1][#2]{#3}}}
  \newcommandtwoopt{\citeyearads}[3][][]%
    {\href{http://adsabs.harvard.edu/abs/#3}
    {\def\hyper@linkstart##1##2{}%
     \let\hyper@linkend\@empty\citeyear[#1][#2]{#3}}}
\makeatother

\begin{document}
\title{Fourier series for eclipses on exoplanet binaries}
\titlerunning{Fourier series for eclipses}
\author{P.M.\ Visser and M.A.\ Mol}
\institute{
Delft Institute of Applied Mathematics, Technical University Delft,
Van Mourik Broekmanweg 6, 2628 XE Delft, The Netherlands \\
\email{p.m.visser@tudelft.nl}
}
\date{Received August 20, 2019\ / Accepted November 15, 2019}
\abstract{
A double planet system or planet binary undergoes eclipses that modify the reflective light curve. In the time domain, the eclipse events are fast and weak. This would make their signal difficult to find and recognize in the phase light curve, even for small inclinations when eclipses happen frequently. However, due to the quasiperiodic nature of the phenomenon, the Fourier transform of the direct reflection signal consists of a double sum of sharp peaks.  These peaks can be resolved for large close binaries and sufficiently long observation times with a star coronagraph.
}
{Eclipses modulate the phase curve, having 
an orbital period $2\pi/\omega$, with a contribution from the relative motion in the binary plane of a period $2\pi/\Omega$.
This leads to a spectral structure with basis frequencies $\omega$ and $\Omega$.
We aim to characterize these spectra.
}
{We studied the regime of short eclipses that occur when the planet radii are small compared to the planet separation. We derived formulas for the peak amplitudes applicable to homogeneous (Lambertian) planet binaries in circular orbit with small inclination.
}
{The effects of an eclipse and of double reflection appear as first- and second-order contributions (in planet radius over separation) in the reflection signal respectively. 
Small peaks appear as observable side bands in the spectrum. Identical structures around $m\Omega$ are characteristic of short-duration eclipses. Deceasing side bands could indicate double reflection between companions.
}
{Fourier analysis of the light curve of non-transiting planets can be used to find planets and their moons. Difficulties in interpreting the structures arise for small planet separation and when there are several moons in mean-motion resonance.
}
\maketitle

\section{Introduction}
The discovery of extensive multi-planet systems around other stars \citep{Lovis2011,Gillon2017,Shallue2018} shows that the Solar System is not unique. Therefore, we may expect exoplanets to also have their own satellites, like the Solar-System planets; Mercury and Venus being exceptions \citep{Namouni2010,Ogihara2012,Barr2016}.
Simulations show that in three-planet systems, two planets often cross orbit and then bond via tidal dissipation, forming a pair called a planet binary \citep{Ochiai2014,Lewis2015}. Knowledge of planet binaries and moons is  important for our understanding of the origins and formation of planetary systems.
A large moon can determine the stability of the planet's spin, generate strong (ocean) tides, and lock its companion into a spin-orbital resonance. These effects in turn influence (geo)physical processes in the crust, oceans, and atmosphere of the planet companion.

Direct imaging of exoplanets is beginning to come within reach with a new generation of ground-based telescopes, that is,
the Thirty Meter Telescope and the Extremely Large Telescope, and two dedicated space telescopes, the James Webb Space Telescope and the Wide Field Infrared Survey Telescope, which have star-occulting coronagraphs with an angular resolution of below $10^{-1}\mathrm{au.}/\mathrm{ly.}$ and contrast of $10^{-6}$ \citep{Boccaletti2004,Krist2007,Douglas2018}. However, the spatial resolution of an exomoon or exoplanet binary requires an angular resolution of much less than a milliarcsecond.
This will not be possible in the foreseeable future, not even for the nearest stars. Therefore,
astronomers will have to rely on a single time-dependent light signal, which is the sum of the light from the parent star, the infrared emission, and the total reflected light from all planets and their moons. Because the phases of the two companions (in orbit around their parent) are the same, their contributions to the modulated radial velocity of the star are equal, as are the contributions to the phase light curve. This makes their individual contribution to the signal indistinguishable. However, there do exist several effects that may reveal the presence of an exomoon:
\begin{inparaenum}[(i)]
\item the photocenter wobble,
\item the Rossiter-MacLaurin effect during a transit, and 
\item the transit timing variation or transit timing duration, which can be detected using the methods of \citet{Kipping2009} and \citet{HellerHippke2016}.
\citet{CabreraSchneider2007,Cabrera2007} proposed the use of
\item the planetary and lunar transits
that occur when the two binary companions become precisely aligned with the observer, and
\item the eclipses that occur for alignments between the planets and the star. 
\end{inparaenum}

During an eclipse, the shadow from the planet or moon temporarily reduces the reflection signal of the companion. This leads to a small reduction in the already weak phase light curve for the short duration of the event. Although small, the eclipses are 
actually the dominant effect from a moon in the reflective phase light curve. The close binary Jupiters found in the simulations by \citet{Ochiai2014} have separations of between four and eight times their radii. In such a system, the shadows would be large and eclipses would happen frequently.

\section{Fourier series}
\begin{figure}
\centering
\resizebox{\hsize}{!}{
\includegraphics{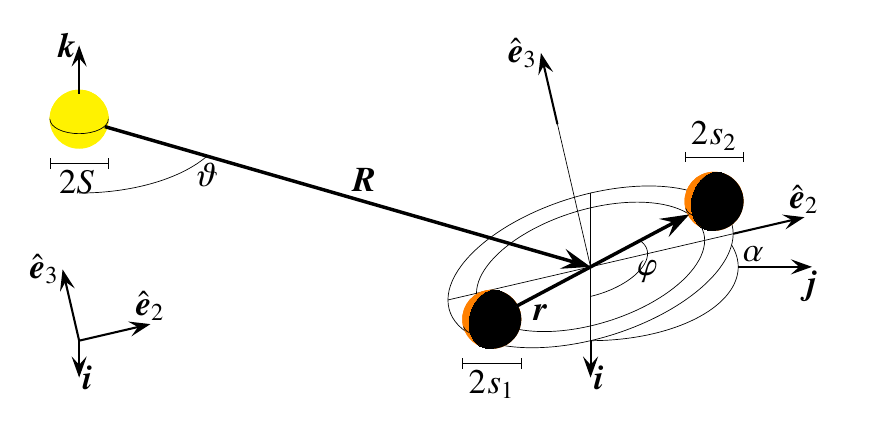}
}
\caption{
\label{Fig1}
Geometry: The central star (yellow) is at the origin of the $\boldsymbol i\boldsymbol j\boldsymbol k$-frame. Vector $\boldsymbol R$ points to the binary barycenter, and the vector $\boldsymbol r$ points from the planet to the moon (orange day sides). The orbital plane of the binary barycenter has normal $\boldsymbol k$. The lunar plane has normal $\hat{\boldsymbol e}_3$ and is inclined by angle $\alpha$. The angles $\vartheta$ and $\varphi$ denote the orbital and lunar phases. The ascending node is in direction $\boldsymbol i$. 
}
\end{figure}

\begin{figure}
\centering
\resizebox{\hsize}{!}{
\includegraphics[width=9cm]{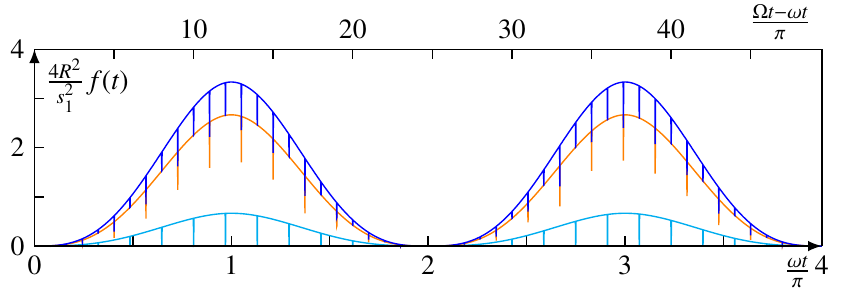}
}
\caption{
\label{Fig2}
Typical light curve in the time domain for a planet binary, as observed edge on. The system has zero inclination and the planet radii have ratio $s_1/s_2=2$. The bottom horizontal axis shows two annual periods, the top horizontal axis shows the lunar periods. Orange: signal from the planet, light blue: signal from the moon, blue: nett signal. The frequencies have ratio $\Omega/\omega=254/19$, as for Earth's approximate Metonic cycle, so that the overall periodicity is actually $19$ orbits.
}
\end{figure}

In all generality, the light signal at a time $t$ from an exoplanet binary is a function of the geometric configuration at that time $t$. There are two phase angles involved: the orbital phase (mean anomaly) $\vartheta$ of the motion of the barycenter and the lunar phase $\varphi$ of the relative motion between the two companions (See Fig.\ \ref{Fig1} and Table.\ \ref{table1}). Therefore, we may write the ideal reflection signal from a planet binary as
\[
f(t) = f(\vartheta,\varphi) , \quad
\vartheta = \omega t , \quad
\varphi = \Omega t
.
\]
We denote with $\omega$, $\Omega$ the mean motions: these are the angular frequencies of the respective barycenter motion and the relative motion. The sidereal month is then equal to $2\pi/\Omega$ and the synodic month is $2\pi/(\Omega-\omega)$. Because the phase angles are periodic variables, $f$ is a double periodic function when considered as a function of $\vartheta$ and $\varphi$, with periods of $2\pi$. Substitution gives a quasiperiodic function of time, with the multi-Fourier series:
\begin{equation}
f(t) = f(\omega t,\Omega t) = \sum_{n=-\infty}^\infty \sum_{m=-\infty}^\infty \mathrm e^{\mathrm i(n\omega+m\Omega)t} f^m_n
.
\label{fourier}
\end{equation}
If the planets are inhomogeneous, the signal can also contain the diurnal period. In fact, the three-body system may have up to 15 basis frequencies, but these include the star's spin as well as very slow precessions. Figure \ref{Fig2} shows an example of a phase curve with the two periods of an exoplanet binary.

\begin{figure}
\centering
\resizebox{\hsize}{!}{
\includegraphics[width=9cm]{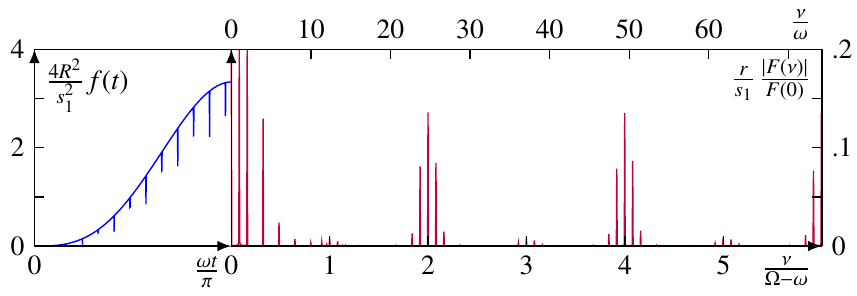}
}
\caption{
\label{Fig3}
Left: Light curve from Fig.\ \ref{Fig2} for one halve orbit. Right: corresponding signal in the Fourier domain. The bottom horizontal axis are in steps of the (fast) lunar frequency. The top horizontal scale are in steps of the (slow) orbital frequency. The side structures that appear due to the eclipses are centered around multiples of $\Omega-\omega$. They are nearly identical copies of each other. Since the eclipse magnitudes of the two types are comparable (and the albedos are equal), the structures at odd $m$ have almost disappeared. Noise would wash out the dips in the time domain and broaden these ideal Fourier peaks.}
\end{figure}

The planets are illuminated by light from the host star that may have a variable intensity $I(t)$. On orbital timescales the noise is caused by star spots and solar-type cycles. Therefore, the observed signal is the product $F(t)=f(t)I(t)$ of the ideal quasiperiodic function (\ref{fourier}) with the intensity of the star. The Fourier transform (truncated to observation duration $T$) of a measured reflection signal from the two planets has the form
\begin{equation}
F_T(\nu) = \sum_{n=-\infty}^\infty \sum_{m=-\infty}^\infty f^m_n I_T(\nu-n\omega-m\Omega)
.
\label{Fnu}
\end{equation}
It has distinct peaks, as shown in Fig.\ \ref{Fig3}. One sees from this expression that every coefficient in Eq. (\ref{fourier}) is the amplitude of a peak in Eq. (\ref{Fnu}). The peak shape $I_T(\nu)$ is the Fourier transform of the intensity spectrum of the source $I(t)$. The peak height is $I_0T^{1/2}$, if we call $I_0$ the average of $I(t)$. Expression (\ref{Fnu}) also shows that noise from the star spills over to the neighboring peaks in the spectrum: stellar noise around $\omega$ or $\Omega$ reduces the visibility of the peaks. Hence, the noise level in the power spectrum of the star at the orbital frequencies determines the visibility of the peaks.

Without coronagraph one receives the nett signal $I(t)+F(t)$ and the spectrum is $I_T(\nu)+F_T(\nu)$.
The coefficients ${f_n}^m$ that arise from an exoplanet are of the order of $N^{1/2}s^2/R^2$ ($s$ and $R$ are the planet and orbital radii, $N$ is the number of observed orbits) and their contribution must exceed the stellar noise in order to be visible. The famous light curves from the hot Jupiters found by \citet{Borucki2009} and \citet{Snellen2009} show that it is possible to observe phase curves without an occultor. Without the transit dip, these stars might have been discarded, while a Fourier transform would have shown huge peaks from the phase variation alone. One would however still need a close-in binary with a short inter-planet distance to detect the effect of an actual eclipse. Blocking the direct starlight with a coronagraph allows measurement of ${f_0}^0$ and eases this severe restriction for the other peaks to the condition $|{f_n}^m|\gg 10^{-3}{f_0}^0$ as was shown in \citet{Visser2015}. In Table \ref{table2} we give estimates of the order of magnitude of the principal peak ${f_1}^0$. The calculation of the (intensity) spectrum $F_T(\nu)$ for a $N$-fold orbit phase light curve allows
\begin{inparaenum}[(i)]
\item separation of individual planet contributions in a multi-planet system \citep{Kane2013},
\item removal of stellar noise, and
\item amplification of the signal by $N^{1/2}$.
\end{inparaenum}
Project Blue \citep{Belikov2015,Morse2018} proposes to make a long and continuous observation of the Alpha-Centauri system with an occultor, obtaining a signal that would be ideal for Fourier analysis.

\section{Peak pattern in the spectrum of binary eclipses}
\label{SecII}
We decompose the reflective light curve of a double planet as:
\begin{equation}
f(t) = f_\mathrm{D}(t) + f_\mathrm{E}(t) + f_\mathrm{S}(t)
,
\label{decomp}
\end{equation}
where D, E, and S indicate the components for direct reflection, eclipses, and secondary reflections. The D term sums two individual planet contributions, where light is reflected off either planet directly towards the observer. It is the pure phase curve with the period $2\pi/\omega$ of the orbit. The E term describes eclipses. This term is negative: it subtracts the contributions of rays that are blocked by the companion. The third S term accounts for rays that are first scattered off one planet in the direction of the companion and are then scattered a second time into the observer direction. 
The measured Fourier spectrum (\ref{Fnu}) has a corresponding decomposition with coefficients of the form
\[
f^m_n = {f_\mathrm{D}}_n \delta^m_0 +  {f_\mathrm{E}}^m_n +  {f_\mathrm{S}}^m_n
.
\]
Here, the Kronecker-delta is introduced to cancel the frequencies $m\Omega$ in the direct component, since these do not occur.

We assume the orbits
of the barycenter and of the relative motion are circular. The longitude angles of the two motions (the true anomalies) are therefore equal to the orbital phases (the mean anomalies). Now we consider eclipses. The planet (number 1) is between the central star and the moon (planet number 2) for $\Omega t=\omega t+2\pi k$, while the moon is between the star and the planet for $\Omega t=\omega t+2\pi k+\pi$ (here $k$ is an integer). We denote the times for these events with
\begin{equation}
t_k = \frac{2\pi k}{\Omega-\omega}
, 
\quad
\bar t_k = \frac{2\pi k+\pi}{\Omega-\omega}
.
\label{tk}
\end{equation}
We consider first the case that the planet is very dark and that the moon is very bright; the albedos are $a_1=0$ and $a_2=1$, so in this case
only the lunar eclipses at $t=t_k$ are observable.
In the summary at the end of this paper, we put the albedo factors back in the equations.
The separation between two (possible) successive eclipse events is the synodic month: $t_{k+1}-t_k=2\pi/(\Omega-\omega)$. The distance $r$ between planet and moon will generally be large compared to the radii, $s_1$ and $s_2$. Because the velocity at which the shadow moves over a surface is equal to $v=|\Omega-\omega|r$, the duration of an eclipse is less than or equal to $(2s_1+2s_2)/v$. Figure 2 shows that the eclipses are of comparatively short duration.

We now approximate the contribution from eclipses to the light curve as a sum of delta functions. In terms of the phase variables we write
\begin{equation}
f_\mathrm{E}(\vartheta,\varphi) =
\frac{v}{r}g(\vartheta)\sum_{k=-\infty}^\infty \delta(\varphi-\vartheta-2\pi k)
.
\label{delta}
\end{equation}
As function of time, using Eq. (\ref{tk}) this becomes
\begin{equation}
f_\mathrm{E}(t) = f_\mathrm{E}(\omega t,\Omega t) =
\sum_{k=-\infty}^\infty g(\omega t_k)\delta(t-t_k) 
.
\label{deltatn}
\end{equation}
The function $g$ can be interpreted as the time integral of an eclipse occurring around phase $\vartheta=\omega t$. It is also periodic in $\vartheta$. The approximation ignores the details of the peaks in the time domain.

It follows from Fourier theory that the continuous Fourier transform of the sum of equally spaced delta peaks (\ref{deltatn}) in the time-domain is a periodic function in the frequency domain, with periodicity $\Omega-\omega$. Hence, we have the identity
\[
\sum_n \sum_m {f_\mathrm{E}}^m_n I_T(\nu-n\omega-m\Omega+\Omega-\omega) = 
\sum_n \sum_m {f_\mathrm{E}}^m_n I_T(\nu-n\omega-m\Omega)
.
\]
For the frequency behavior to be periodic in $\nu$ by $\Omega-\omega$, the coefficients must be related by ${f_\mathrm{E}}^{m+1}_{n-1} = {f_\mathrm{E}}^m_n$. However, this implies ${f_\mathrm{E}}^m_n = {f_\mathrm{E}}^0_{n+m} $. The overall Fourier transform (\ref{Fnu}) of a detected signal, neglecting double reflections for the moment, is therefore
\[
F_T(\nu) = \sum_n {f_\mathrm{D}}_n I_T(\nu-n\omega) + \sum_n\sum_m {f_\mathrm{E}}^0_{n+m} I_T(\nu-n\omega-m\Omega)
,
\]
and is characterized by the two sets of coefficients ${f_\mathrm{D}}_n$, ${f_\mathrm{E}}^0_n$. The effect of reflection between companions is discussed in Sect.\ \ref{SecVI}.

Figure \ref{Fig3} shows a spectrum of an eclipsing binary with two eclipses every month: one lunar and one planetary eclipse. The peaks at the frequencies $n\omega$ that correspond to the annual motion are found near the origin. These peaks have values ${f_0}^0$, ${f_1}^0$, ${f_2}^0$, and ${f_3}^0$, and so on. They arise predominantly from direct reflection and have almost
the same values as the spectrum of one planet. The structures around $\nu=m\Omega$ for $m\neq 0$ are due to eclipses. They have peak amplitudes like ${f_0}^m$, ${f_1}^m$, ${f_2}^m$, and ${f_3}^m$. One has to compare these with the first side structure around $\nu=\Omega$. For a dark planet and a bright moon (or for a dark moon and a bright planet), the side bands at $m=1$ and $m=2$ are copies of one another,
because the peak values are related by ${f_{-1}}^1={f_{-2}}^2$, ${f_0}^1={f_{-1}}^2$, ${f_1}^1={f_0}^2$, and so on. If both companions are bright there is destructive interference at the odd values for $m$. Because of the doubling of eclipses, the structure in the Fourier transform repeats after $2\Omega-2\omega$. This is the case in Fig.\ \ref{Fig3}, where the even and odd side bands are comparable. This pattern of identical copies is repeated as long as $|m|\ll r/s_1$. The motion of a moon and planet around each other thus gives rise to side bands in the spectrum at the frequency $\Omega$, with a smaller structure of peaks separated by the $\omega$ of the annual motion. The peak ${f_{-1}}^1$ at $\Omega-\omega$ is the average of the (integrated) eclipse dips and should be negative (for a suitable choice of phase) and is the largest peak in the first side band. Retrograde relative motion has a negative $\Omega$.

The short duration of the eclipse events results in frequency side bands that are approximately identical. An alternative derivation of this elementary result is the following. The Fourier coefficients of (\ref{fourier}) can be expressed as the double integral:
\begin{equation}
f^m_n = \frac{1}{(2\pi)^2}\int\limits_0^{2\pi}\mathrm{d}\vartheta \int\limits_0^{2\pi}\mathrm{d}\varphi\, \mathrm e^{-\mathrm in\vartheta-\mathrm im\varphi} f(\vartheta,\varphi)
.
\label{fnm}
\end{equation}
We consider short eclipse duration. Consequently,
the integrand in (\ref{fnm}) from the eclipse contribution $f_\mathrm{E}$ is only nonzero for times $t$ near (\ref{tk}). This implies that the phase difference $\vartheta-\varphi$ is near an integer multiple of $2\pi$; see identity (\ref{delta}). We may therefore replace the exponent $e^{-\mathrm in\vartheta-\mathrm im\varphi}$ in the integrand with $e^{-\mathrm in\vartheta-\mathrm im\vartheta}$, so that a good approximation is obtained: 
\[
{f_\mathrm{E}}^m_n = \frac{1}{(2\pi)^2}\int\limits_0^{2\pi}\mathrm{d}\vartheta\, \mathrm e^{-\mathrm i(n+m)\vartheta} \int\limits_0^{2\pi}\mathrm{d}\varphi\,  f_\mathrm{E}(\vartheta,\varphi) = {f_\mathrm{E}}^0_{n+m}
. 
\]
By integrating (\ref{delta}) over one period of the independent variable $\varphi$, one obtains the $\vartheta$-periodic function $g(\vartheta):$
\[
\frac{r}{v} \int\limits_0^{2\pi} f_{\mathrm E}(\vartheta,\varphi)\mathrm d\varphi = g(\vartheta) =
\sum_{n=-\infty}^\infty g_n \mathrm e^{\mathrm in\vartheta}
,
\]
which has Fourier coefficients $g_n$. After using these results again in the combination of (\ref{delta}) with (\ref{fnm}), one obtains
\begin{equation}
{f_\mathrm{E}}_{n+m}^0 =
\frac{v}{(2\pi)^2r}\int\limits_0^{2\pi}\mathrm{d}\vartheta\, \mathrm e^{-\mathrm i(n+m)\vartheta} g(\vartheta) =
\frac{v}{2\pi r} g_{n+m}
.
\label{pattern}
\end{equation}
The identical side-band structures in the spectrum are given by the coefficients of the periodic function $g$ describing the time-integrated dips due to eclipses as a function of orbital phase, multiplied by the number of eclipses per unit time.

\section{Numerical and observational implementation}
\label{SecIII}
In this paper we implicitly assume that $\Omega$ and $\omega$ do not have a simple ratio, so that $f$ is quasiperiodic.
However, if the ratio of the frequencies is simple, that is,\  $\Omega/\omega=h/k$ with $h$, $k$ small coprime integers, the light curve is purely periodic, with an overall period $2\pi k/\omega$. This situation would correspond to a peculiar type of orbital resonance. Due to the fact that now $n\omega+m\Omega=(nk+mh)\omega/k$, many pairs $n,m$ correspond to the same frequency, each contributing to the same spectral peak. It is no longer possible to find the individual coefficients ${f_n}^m$ from the Fourier-transformed light curve.

We are interested in the case where $\Omega/\omega$ is not a simple ratio. However, in order to numerically simulate the system, we have to choose a smallest time-step $\mathrm dt$ and a total integration time $T$. Because we want our theoretical Fourier peaks to be sharp, we require an overall periodicity. The most accurate approach is to use two coprime multiples of $\mathrm dt$ for the two periods. This also allows the eclipse maxima $t_k$, $\bar t_k$ and the maxima for the orbital phase to occur at exact data points. The fast Fourier transform cannot be used, since our domain size is not a power of two.

The Babylonians and ancient greek astronomers also wanted to approximate the ratio of the duration of a year to that of a month on Earth and they used $\Omega/\omega=1+235/19$. The overall period of $19$ years is called the Metonic cycle; see \citet{Pannekoek1947}. In the numerical calculations for all figures in this paper, we also used this ratio. Obviously quasiperiodic motion is better approximated as the numbers $h$, $k$ become larger.

For analysis of observational data we encounter the same problem because there is also a smallest observational time resolution $\mathrm dt$ and a total observation duration $T$. If $\omega$ is known in advance, it may help to consider an integral number of orbits, but the peaks $m\Omega$ will be displaced due to the random cutoff, roughly by $\omega N^{-1/2}$. On the other hand, it could be that the relevant frequencies only appear after the Fourier transform has been made.
In that situation, comparing $F_T$ with $I_T$ 
seems best.

For planets very close to an M-type star, it may be possible to obtain an uninterrupted signal for several orbits. Fourier peaks may be found even without spatially isolating the planets  (as individual points) from the star. For planets near a bright G-type star, the direct light must be blocked with a coronagraph \citep{Cash2006,Mawet2010}. When continuous observations are not available due to the length of the period, several short-duration observations along the orbit may be sufficient. For example, one could make four separate observations during intervals of equal length $\Delta T$, along points that are $90^\circ$ apart in orbital phase. If $\Delta T$ is several lunar periods but still small compared to the annual period, that is,\ $\omega\ll (\Delta T)^{-1}\ll\Omega$, an approximation for the peaks in Equation (\ref{Fnu}) is
\[
f^m_n \approx \frac{1}{4\Delta T I_0} \sum_{k=0}^3 \mathrm i^{-nk} \int_{\frac{\pi k}{2\omega}-\frac{\Delta T}{2}}^{\frac{\pi k}{2\omega}+\frac{\Delta T}{2}} \mathrm e^{-\mathrm im\Omega t}F(t) \mathrm dt
, \quad n=0,1,2,3
.
\]
One could neglect $f^m_4$ and higher $n$ because these coefficients are small and decay fast with $n$. This method is analogous to combining several telescopes in astronomical interferometry.
Now, peak height of the side bands scales as $\Omega\Delta T$, not as $\Omega/\omega$.

\begin{table}
\caption{Parameters used in modeling}
\label{table1}      
\centering
\begin{tabular}{l|l}
\hline
\hline
symbol & quantity \\
\hline
$t$ & time \\
$t_k$, $\bar t_k$ & time of lunar, planetary eclipse \\
$T$ & observation duration \\
$L$ & distance to the observer on Earth \\
$S$ & star radius \\
$\alpha$ & binary inclination angle, w.r.t.\ orbital plane \\
$s_1$, $s_2$ & planet, moon radius \\
$a_1$, $a_2$ & planet, moon albedo \\
$\boldsymbol R_1$, $\boldsymbol R_2$ & planet, moon position vector \\
$\boldsymbol R=\hat{\boldsymbol R}R$ & binary c.m.-position vector \\
$\boldsymbol r=\hat{\boldsymbol r}r$ & reduced position vector \\
$\boldsymbol s=\hat{\boldsymbol s}s$ & arbitrary planet surface vector \\
\hline
$\theta$ & azimuth angle \\
$z$ & vertical coordinate w.r.t.\ planet center \\
$l$ & vertical displacement of shadow center \\
$v$ & shadow velocity \\
$\tau(l)$ & eclipse duration \\
$\Upsilon(l)$ & eclipse magnitude \\
\hline
$\nu$ & continuous frequency variable \\
$\omega$ & barycenter angular frequency, mean motion \\
$\vartheta=\omega t$ & barycenter phase angle, mean anomaly \\
$\Omega$ & lunar orbital frequency \\
$\varphi=\Omega t$ & lunar phase \\
\hline
$\hat{\boldsymbol o}$ & observation direction \\
$\theta_\mathrm{o}$ & observation inclination, polar angle \\
$\phi_\mathrm{o}$ & phase at inferior conjunction, azimuth \\
$I(t)$ & variable star luminosity \\
$I_T(\nu)$ & Fourier transform of star luminosity \\
$I_0$ & average star luminosity \\
$F(t)$ & observed reflection signal \\
$F_T(\nu)$ & Fourier transform of reflection signal \\
$f(t)=f(\vartheta,\varphi)$ & total reflected light curve \\
$f_n^m$ & Fourier coefficient of light curve \\
$g(\vartheta)$ & time-integrated dip for lunar eclipse \\
$g_n$ & Fourier coefficient of  $g$ \\
$h(\vartheta)$ & single-planet phase light curve \\
$h_n$ & Fourier coefficient of $h$ \\
\hline
$\boldsymbol i$, $\boldsymbol j$, $\boldsymbol k$,  & basis vectors for orbital plane \\
$\hat{\boldsymbol e}_1$, $\hat{\boldsymbol e}_2$, $\hat{\boldsymbol e}_3$ & basis vectors for lunar plane \\
$N=\omega T/2\pi$ & number of orbits \\
$k$ & eclipse index number \\
$n$, $m$ & integer indices \\
\hline
\hline
\end{tabular}
\tablefoot{Symbol and significance of the physical quantities used.}
\end{table}

\begin{figure}
\centering
\resizebox{\hsize}{!}{
\includegraphics{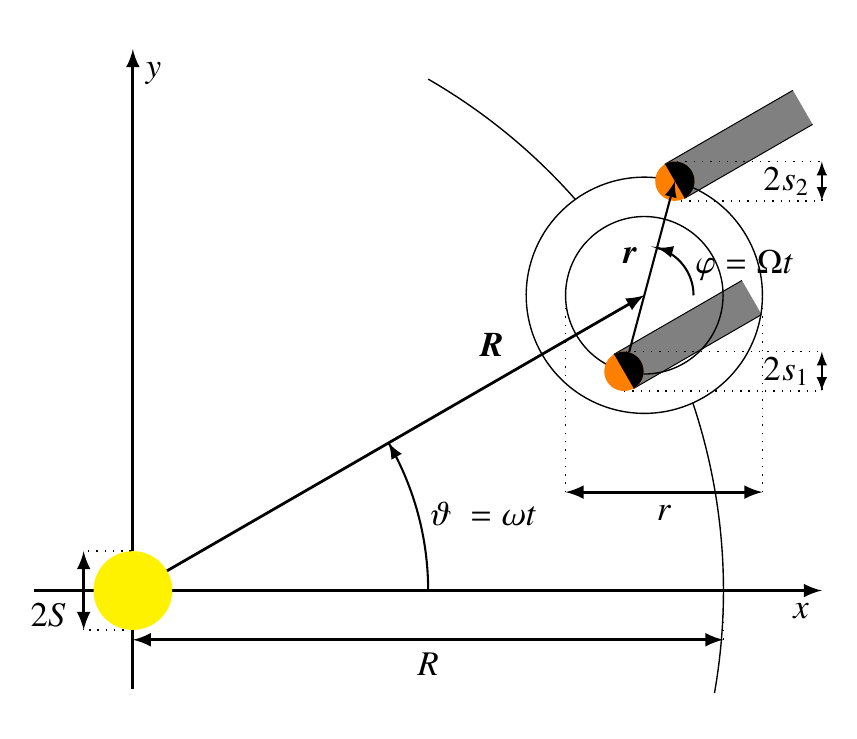}
}
\caption{
\label{Fig4}
Top view of the system. For a lunar eclipse, the moon is in the shadow of the planet, and the orbital phases must be equal: $\vartheta=\varphi$ modulo $2\pi$. For a planetary eclipse, with the planet in the shadow of the moon, the orbital phases are opposite: $\vartheta=\varphi+\pi$ modulo $2\pi$.
}
\end{figure}

\begin{figure}[t]
\centering
\resizebox{\hsize}{!}{
\includegraphics{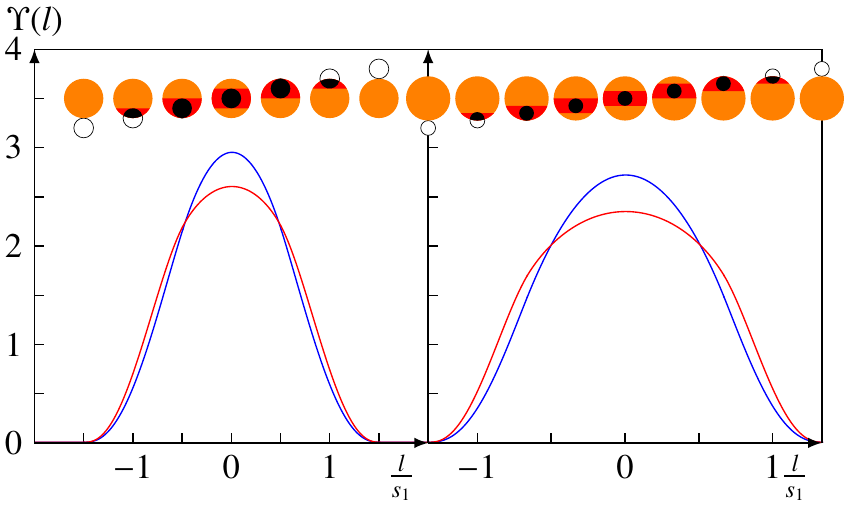}
}
\caption{
\label{Fig5}
Eclipse magnitude $\Upsilon$, defined in (\ref{defM}), vs. displacement $l$ of the shadow trace, for radii $s_1=2s_2$ (left) and $s_1=3s_2$ (right). The visuals show the viewpoint from the star: the eclipses (both types) trace out a nearly horizontal band. Blue and red: $\Upsilon$ and $\bar\Upsilon$ for lunar and planetary eclipses, respectively. For $|l|=s_1\pm s_2$, the shadow cylinder of one partner grazes its companion (and the projected disks touch). There, the third derivative is infinite, resulting in a tail in the Fourier spectrum.
When $\Upsilon(l_k)$ is multiplied with the phase curve $h(\vartheta_k-\phi_\mathrm{o})$, one obtains $|g(\vartheta_k)|$ which equals the integrated dip in the light curve for edge-on observation of an eclipse.
}
\end{figure}

\section{Eclipses in the time domain}
\label{SecIV}
\subsection{Description of planet binary}
Let the positions of the planet and its moon be the vectors $\boldsymbol R_1$ and $\boldsymbol R_2$ and let $\boldsymbol R$ be the coordinate vector from the star to the barycenter of the binary. The difference vector $\boldsymbol r=\boldsymbol R_2-\boldsymbol R_1$ is the relative coordinate (pointing from the larger planet to the smaller moon). The position vectors are given by
\begin{align}
\boldsymbol R(t) &= \hat{\boldsymbol R}(t)R = (\boldsymbol i\cos\omega t+\boldsymbol j\sin\omega t)R
,
\label{circles} \\
\boldsymbol r(t) &= \hat{\boldsymbol r}(t)r = (\boldsymbol i\cos\Omega t+\hat{\boldsymbol e}_2\sin\Omega t)r
.
\nonumber
\end{align}
The ascending node is in direction $\boldsymbol i$ if the lunar plane is inclined with respect to the orbital plane. See Fig.\ \ref{Fig1} for the geometry.
For the orbits to be stable, the planet separation distance $r$ must be smaller than $R$. We assume that the bodies are spheres, with radii $s_1$, $s_2$ that are small compared to $r$, and that the stellar radius $S$ is small compared to the distance $R$ 
between the star and the barycenter of the binary. We thus consider the regime where
\begin{equation}
s_2 \leq s_1 \ll r \ll R , \quad S \ll R \ll L
.
\label{approx}
\end{equation}
As $L$ is the distance of the system to Earth, the system could be spatially resolved from the star for inner working angles below $R/L$. One requires an unfeasible angular resolution below $r/L$ for separation of the planet and moon.

Now, when one planet moves between the star and its companion, an eclipse does occur when the shadow of the planet falls onto the companion. One instant of a lunar eclipse is found at $t=0$, because at that 
moment $\boldsymbol R=R\boldsymbol i$ and $\boldsymbol r=r\boldsymbol i$ and therefore $\vartheta=\varphi=0$.
The eclipse is then at maximum, with the moon completely in the shadow of the planet.

Because $S\ll R$, light rays are almost parallel when they hit the surface of the  planet. In that case, the distance between the moon center and the axis of the shadow cylinder is $|\hat{\boldsymbol R}\times \boldsymbol r|$. We introduce the displacement $l=l_k$ as the minimal value of this distance. This is simply the minimal distance between the disk centers when the planets are projected on the plane normal to $\hat{\boldsymbol R}$, that is,\ from the viewpoint of the star.
In general, $|\hat{\boldsymbol R}\times \boldsymbol r|$ is minimal for
\[
\frac{\mathrm{d}(\boldsymbol R\bullet\boldsymbol r)}{\mathrm{d}t} = 0
.
\]
The minima indeed occur at the times given by (\ref{tk}). 
We refer to an eclipse as \emph{complete} if the moon gets inside the shadow cylinder of the larger planet or if the planet fully intercepts the shadow cylinder of the moon. From the viewpoint of the central star, the disks of the two bodies overlap each other. In the parallel-ray approximation, the complete eclipses arise for
\begin{equation}
l_k < s_1 - s_2
.
\label{complete}
\end{equation}
We refer to an eclipse as \emph{partial}  when the shadow cylinder of one planet just intersects the other planet. The condition is
\begin{equation}
s_1 - s_2 < l_k < s_1 + s_2
.
\label{partial}
\end{equation}
There is no eclipse at $t_k$ if $s_1+s_2<l_k$. Our distinction between complete and partial eclipse is the same as for lunar eclipses on Earth. However, in our approximation $S\ll R$, on the surface, the penumbra is negligible compared to the umbra \citep{Link1969}. Of course, the distinction between total and partial \emph{solar eclipse} on Earth usually refers to different observer locations.

Exoplanet eclipses must occur frequently in order that they may be recognized. Therefore, inclination of the lunar plane with respect to the barycenter plane must be sufficiently small. We assume $0\leq\alpha\ll 1$ and approximate the unit basis vectors in (\ref{circles}) by
\[
\hat{\boldsymbol e}_2 = \boldsymbol j + \alpha\boldsymbol k , \quad 
\hat{\boldsymbol e}_3 = \boldsymbol k - \alpha\boldsymbol j .
\]
The displacement for eclipse at time $t_k$ is now
\begin{equation}
\pm l_k = {-r}\hat{\boldsymbol R}(t_k)\bullet\hat{\boldsymbol e}_3 = \boldsymbol r(t_k)\bullet\boldsymbol k = \alpha r \sin\omega t_k = \alpha r \sin\Omega t_k
.
\label{ln}
\end{equation}
The eclipses can occur frequently, but not necessarily every month. We consider three cases:
\begin{inparaenum}[(i)]
\item
Complete eclipses occur every month when $l_k$ is always less than the difference in the planet radii. Because $\pm l_k$ oscillates between $-\alpha r$ and $\alpha r$, this is the case where $\alpha r<s_1-s_2$. The Galilean satellites Io, Europa, and  Ganymede fall in this case.
\item Eclipses occur every month but are sometimes {partial} in cases where $s_1-s_2<\alpha r<s_1+s_2$.
\item Eclipses do not always occur for $s_1+s_2<\alpha r$, like for Jupiter-Callisto and the Saturn-Titan, Pluto-Charon, and Earth-Moon systems.
\end{inparaenum}

\subsection{Eclipse duration and eclipse magnitude}
As an eclipse happens on the moon (or the planet), the shadow of one body crosses the surface of its companion. It will be convenient to parametrize the surface vector $\boldsymbol s$ in cylindrical coordinates $\theta$, $z$, as:
\[
\boldsymbol s = (\boldsymbol i\cos\theta + \boldsymbol j\sin\theta)\sqrt{s^2-z^2} + \boldsymbol kz
.
\]
The surface element in cylindrical coordinates is $\mathrm{d}^2A= s\mathrm{d}\theta \mathrm{d}z$. From the viewpoint of the star, this shadow is a disk of radius $s_1$ (or $s_2$). The planet shadow traces out a nearly horizontal band along the surface; see insets in Fig.\ \ref{Fig5}. This band for $z$ is the intersection $[z_-,z_+]=[-s_2,s_2]\cap[l-s_1,l+s_1]$, or
\[
z_- = \mathrm{max}(-s_2,l-s_1)
< z <
z_+ = 
\mathrm{min}(s_2,l+s_1).
\]
The global eclipse duration, which is the time between first contact (of the shadow cylinders) and last contact, is equal to
\[
\frac{2}{v}\sqrt{(s_1+s_2)^2-l^2}
.
\]
The local duration of the eclipse for one fixed point on the moon (in a nonrotating frame) is the cord length at $z$ divided by the shadow velocity:
\[
\tau(z) = \frac{2}{v}\sqrt{{s_1}^2-(z-l)^2}
.
\]
The following definition of a {"magnitude"}, as an average of the eclipse duration as a function of the displacement $l,$ will prove useful: $\Upsilon(l) = $
\begin{equation}
\frac{\displaystyle\int\limits_{z_-}^{z_+}\mathrm{d}z\, ({s_2}^2-z^2)\tau(z)}{\displaystyle\int\limits_{-s_2}^{s_2}\mathrm{d}z\, ({s_2}^2-z^2)}
= \frac{3}{2} \int\limits_{z_-}^{z_+}\frac{\mathrm{d}z}{v}\, \bigg( 1-\frac{z^2}{{s_2}^2}\bigg) \frac{\sqrt{{s_1}^2-(z-l)^2}}{s_2}
.
\label{defM}
\end{equation}
The corresponding eclipse magnitude $\bar \Upsilon(l)$ is of the same form as  Eq. (\ref{defM}) but with $s_1$ and $s_2$ interchanged. Graphs of $\Upsilon$ and $\bar\Upsilon$ are plotted in Fig.\ \ref{Fig5}.
Although the areas of intersection of the projected disks for the two types are equal, the magnitude of the lunar eclipse is only slightly different from the magnitude of the planetary eclipse.

\subsection{General phase light curve}
We now derive the standard phase light curve and the correction due to eclipses. The light output from the star that is directly intercepted by our telescope (without occultor) is equal to the solid angle fraction of the total luminosity $I_0\mathrm{d}^2\hat{\boldsymbol o}/4\pi$. Each element $\hat{\boldsymbol s} \mathrm{d}^2 A$ of planet surface (with unit albedo) that intercepts starlight reflects the following  luminosity into our telescope:
\begin{equation}
I_0\frac{(-\hat{\boldsymbol R}\bullet\hat{\boldsymbol s})\mathrm{d}^2A}{4\pi R^2} \frac{(\hat{\boldsymbol s}\bullet\hat{\boldsymbol o})\mathrm{d}^2\hat{\boldsymbol o}}{\pi}
.
\label{element}
\end{equation}
Here we use Lambert's cosine law for reflection. The second factor for the reflected ray is properly normalized, because
\[
\int\int_{\hat{\boldsymbol s}\bullet\hat{\boldsymbol o}>0}\mathrm{d}^2\hat{\boldsymbol o} \frac{(\hat{\boldsymbol s}\bullet\hat{\boldsymbol o})}{\pi}  = 1
.
\]
The reflection signal is obtained by integrating (\ref{element}). The light flux from the planet and from the star are both proportional to the small solid angle $\mathrm{d}^2\hat{\boldsymbol o}$. Because we consider the relative signal, we factor out $I_0\mathrm{d}^2\hat{\boldsymbol o}/4\pi$. The ideal phase light curve from a Lambertian moon orbiting a dark planet is found to be
\begin{align}
f_\mathrm{D}(t) &= \frac{1}{\pi R^2} \int\int_{\rightmoon} \mathrm{d}^2A\, ({-\hat{\boldsymbol R}(t)}\bullet\hat{\boldsymbol s})(\hat{\boldsymbol s}\bullet\hat{\boldsymbol o})
,
\label{fD}
\\
f_\mathrm{E}(t) &= 
\frac{-1}{\pi R^2} \int\int_{\astrosun}
\mathrm{d}^2A\, ({-\hat{\boldsymbol R}(t)}\bullet\hat{\boldsymbol s})(\hat{\boldsymbol s}\bullet\hat{\boldsymbol o})
.
\label{fE}
\end{align}
The respective integration domains are
\begin{align}
\rightmoon &= \Big\{ \boldsymbol s=s_2\hat{\boldsymbol s} \Big| (-\hat{\boldsymbol R}\bullet\hat{\boldsymbol s}) > 0 , \quad (\hat{\boldsymbol s}\bullet\hat{\boldsymbol o}) > 0 \Big\}
,
\label{lune}
\\
\astrosun &= \Big\{ \boldsymbol s=s_2\hat{\boldsymbol s} \Big| (\boldsymbol R\bullet\boldsymbol r)>0 , \quad |\hat{\boldsymbol R}\times (\boldsymbol r+s_2\hat{\boldsymbol s})| < s_1 \Big\} \cap \rightmoon
.
\label{sun}
\end{align}
For the direct signal, $f_\mathrm{D}$, one must integrate over the spherical lune $\rightmoon$ defined as the intersection of the illumination with the visibility. The (negative) surface integral for the eclipses, $f_\mathrm{E}$, is over the shadow region $\astrosun$ cast by the planet onto the moon. The first condition in $\astrosun$ states that the vector $\boldsymbol r$ points away from the star: the planet is nearest to the star. The second condition states that the distance between the point on the surface of the moon to the axis $\boldsymbol R$ is less than $s_1$. These are the points in the shadow (cylinder) of the planet.

\subsection{Face-on observation}
\begin{figure}
\centering
\resizebox{\hsize}{!}{
\includegraphics[width=9cm]{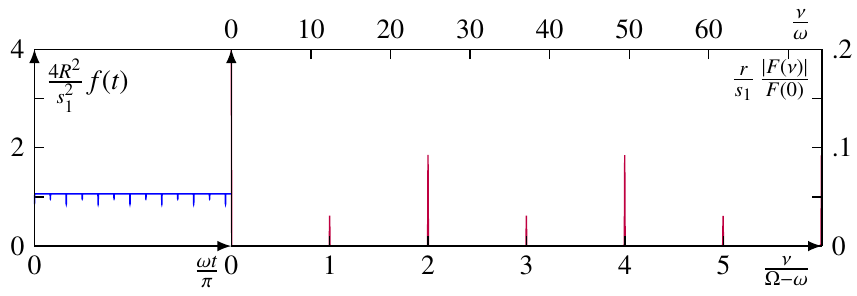}
}
\caption{
Numerical results for the simplest case: face-on observation of a binary with zero inclination. Shown are the signal and the Fourier transform (left and right respectively) for $s_1/s_2=2$ with axes as in Fig.\ \ref{Fig3}. The signal has the lunar periodicity only. In contrast to the edge-on signal, the peak magnitudes for the planet and lunar eclipse differ if $s_1\neq s_2$, meaning that the peaks at odd $m$ are not small as compared to even $m$.
}
\label{Fig6}
\end{figure}

\begin{figure}[t]
\centering
\resizebox{\hsize}{!}{
\includegraphics{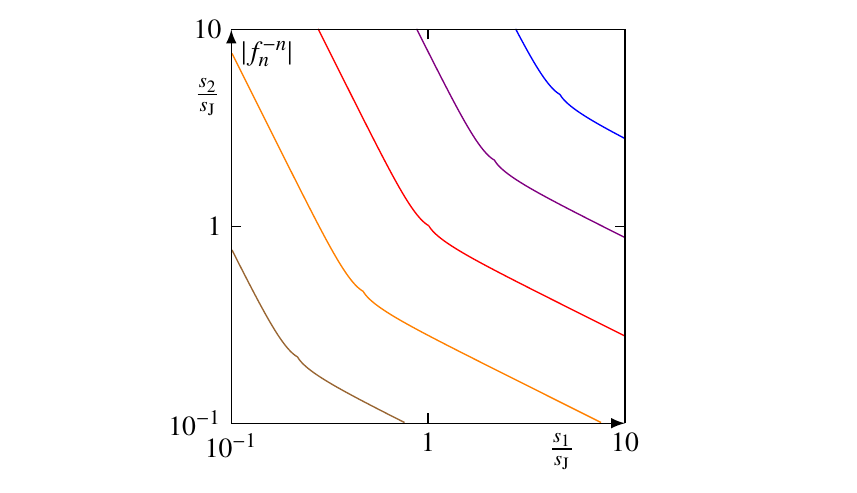}
}
\caption{
Eclipse-peak height for a double planet without inclination, when observed face on, as a function of the two planet radii, from formula\, (\ref{An}). The radius $s_\mathrm{J}$ is of Jupiter and the level curves are for the values $8\pi R^2r|{f_n}^{-n}|/s_\mathrm{J}^3=10^{-2},10^{-1},\ldots, 10^2$ (brown to blue). 
}
\label{Fig7}
\end{figure}

Let us start with the special case of face-on observation. For $\hat{\boldsymbol o}=\boldsymbol k$, there are no observable phases, as the lune is permanently a quarter sphere. However, when the shadow falls on the northern hemisphere, the eclipses are visible. For the calculation of (\ref{fD})-(\ref{fE}) we require
\[
(-\hat{\boldsymbol R}\bullet\hat{\boldsymbol s}) = \cos(\theta-\omega t-\pi)\frac{\sqrt{{s_2}^2-z^2}}{s_2}
, \quad
(\hat{\boldsymbol s}\bullet\hat{\boldsymbol o}) = \frac{z}{s_2}
.
\]
After integration over $z$ in $[0,s_2]$ and over $\theta$, we find
\[
f_\mathrm{D}(t) = \frac{2{s_2}^2}{3R^2}
, \quad
f_\mathrm{E}(t) = \frac{-1}{R^2} \sum_{k=-\infty}^\infty \delta(t-t_k) \int\limits_{z_-}^{z_+}\mathrm dz\, \frac{z\sqrt{{s_2}^2-z^2}}{s_2} \tau(z)
.
\]
The integration interval for $z$ is the intersection of the eclipse band with the northern hemisphere: $[z_-,z_+]=[l-s_1,l+s_1]\cap [0,s_2]$. The boundary points are then
\[
z_- = \mathrm{max}(0,l-s_1)
< z <
z_+ = 
\mathrm{min}(s_2,l+s_1)
.
\]
The simplest case of $\alpha=0$, $l=0$ has recurring eclipses every month. The signal in time and Fourier domain is plotted in Fig.\ \ref{Fig6} for $a_1=a_2=1$. Equation (\ref{pattern}) for the lunar eclipses implies the relation ${f_\mathrm{E}}^m_n = g_0\delta_0^{n+m} v/2\pi r$. The peak values are given by
\begin{equation}
f^{-n}_{\mathrm{E}n} = \frac{-1}{8\pi R^2r} 
\Bigg[ \frac{(s_1^2+s_2^2)s_1}{2} + \frac{(s_1^2-s_2^2)^2}{4s_2}\log\frac{s_1-s_2}{s_1+s_2}
\Bigg]
.
\label{An}
\end{equation}
This is plotted as a function of the radii in Fig.\ \ref{Fig7}. Although the depth of the dips in the time domain can equal the intensity of the normal phase curve (namely for complete eclipses), a dip may be difficult to find due to the short duration of an event. The value of $g_0$ decreases with decreasing $\tau$, but the ${f_n}^{-n}$ remain constant. This is because for increasing $\Omega$, eclipses occur more frequently, and the Fourier transform adds up all events in one peak.

For the planetary eclipses, we obtain the same expression (\ref{An}), except for an overall factor $s_2/s_1$ and with $t_k$ replaced by $\bar t_k$. For face-on view, the planetary eclipses are weaker than the lunar eclipses, because the shadow domain on the planet is flatter than on the moon, and therefore it appears thinner as the view is from the side.

\subsection{Edge-on observation}
Now we consider the case of observation as nearly edge-on to the orbital plane. The condition for observer inclination angle $\theta_\mathrm{o}$ will be $0\leq \frac{\pi}{2}-\theta_\mathrm{o}\ll 1$. Here, $\theta_\mathrm{o}$ is preferably slightly less that 90 degrees, so that the planet does not move behind the coronagraph or behind or in front of the star. The observer direction is approximated by
\[
\hat{\boldsymbol o} = \boldsymbol i\cos\phi_\mathrm{o} + \boldsymbol j\sin\phi_\mathrm{o} + \boldsymbol k(\tfrac{\pi}{2}-\theta_\mathrm{o})
.
\]
Conditions for observer inclination and azimuth $\theta_\mathrm{o}$, $\phi_\mathrm{o}$ for obtaining uninterrupted phase curves, also prohibiting transits and occultations of the planets (where both planets and the observer are aligned), are given in Appendix \ref{AppA}. 

To facilitate the integration over the spherical lune (\ref{lune}), one often introduces the positive part $c(\theta)=\mathrm{max}(0,\cos\theta),$ which is equal to $\cos\theta$ when positive, and zero otherwise \citep{Cowan2008,Cowan2013,Fujii2012}. These can now replace the dot product in expressions (\ref{fD})-(\ref{fE}) and we can extend the azimuth integration to the full range. This gives
\begin{align}
f_\mathrm{D}(t) &= \frac{1}{\pi R^2} \int\limits_{-{s_2}}^{s_2}\mathrm dz\, \frac{{s_2}^2-z^2}{s_2} \int\limits_0^{2\pi}\mathrm d\theta\, c(\theta-\omega t-\pi)c(\theta-\phi_\mathrm{o})
,
\label{convolution} \\
f_\mathrm{E}(t) &=
\label{eclipses}
\\
\frac{-1}{\pi R^2} & \sum_n \delta(t-t_n) \int\limits_{z_-}^{z_+}\mathrm dz\, \frac{{s_2}^2-z^2}{s_2} \tau(z) \int\limits_0^{2\pi}\mathrm d\theta\, c(\theta-\omega t-\pi)c(\theta-\phi_\mathrm{o})
.
\nonumber
\end{align}
Evaluating the integral for the direct signal gives the result
\begin{equation}
f_\mathrm{D}(t) = \frac{{s_2}^2}{4R^2} h(\omega t-\phi_\mathrm{o})
, \quad
h(\theta) = \frac{8\sin|\theta|-8|\theta|\cos\theta}{3\pi} 
.
\label{fDresult}
\end{equation}
Here we used the normalized phase curve $h$ for a single planet from the paper of \citet{vanHulst1980}, for phase angles $\theta=\omega t-\phi_\mathrm{o}$ in the interval $[-\pi,\pi]$. The factor ${s_2}^2/4R^2$ in (\ref{fDresult}) is the fraction of intercepted light. Evaluating the integrals in Eq.\ (\ref{eclipses}) for the eclipses gives the result
\begin{equation}
f_\mathrm{E}(t) = -\sum_k \delta(t-t_k) \Upsilon(l_k) f_\mathrm{D}(t_k)
.
\label{fEresult}
\end{equation}
Comparing this with the general Equation (\ref{deltatn}), the contribution for an eclipse can be read off: $g(\vartheta_k)=-\Upsilon(l_k)f_\mathrm{D}(t_k)$, the product of the eclipse magnitude (\ref{defM}) with the value of the pure phase curve at the time of the event.

\section{Eclipses in the Fourier domain}
\label{SecV}
\subsection{Coefficients for individual planets}
Because the function $c$ is periodic modulo $2\pi$, it has a Fourier series. This is
\begin{equation}
c(\theta) =
\sum_{n=-\infty}^\infty \mathrm e^{\mathrm in\theta}c_n =
\frac{\mathrm e^{\mathrm i\theta}+\mathrm e^{-\mathrm i\theta}}{4} + \frac{1}{\pi} \sum_{\substack{n \\ \mathrm{even}}} \frac{(-1)^{n/2}\mathrm e^{\mathrm in\theta}}{1-n^2}
.
\label{cn}
\end{equation}
Now we substitute this in (\ref{convolution}). Since these integrals are convolutions of $c$ with itself, the coefficient for the $f_\mathrm{D}$ is essentially the square of $c_n$. We find
\begin{equation}
f_\mathrm{D}(t) = 
\frac{s^2}{4R^2} \sum_{n=-\infty}^\infty \mathrm e^{\mathrm in(\omega t-\phi_\mathrm{o})}h_n =
\frac{8s^2}{3R^2} \sum_{n=-\infty}^\infty \mathrm e^{\mathrm in(\omega t-\phi_\mathrm{o})}(-1)^n c_n^2
,
\label{hIdef}
\end{equation}
and
\begin{equation}
h(\theta) = \sum_{n=-\infty}^\infty \mathrm e^{\mathrm in\theta}h_n =
\frac{-2\mathrm e^{\mathrm i\theta}-2\mathrm e^{-\mathrm i\theta}}{3} + \frac{32}{3\pi^2}\sum_{\substack{n \\ \mathrm{even}}} \frac{\mathrm e^{\mathrm in\theta}}{(1-n^2)^2}
.
\label{hn}
\end{equation}
Odd-numbered coefficients, except $h_1$ and $h_{-1}$, are zero. Here,  $h_1$ and $h_{-1}$ are negative because for $\omega t=\phi_\mathrm{o}$ one has an inferior conjunction (binary between star and observer) and then the signal is minimal. The peaks with values ${f_n}^0\approx f_{\mathrm D n}=({s_1}^2+{s_2}^2)h_n/4R^2$ are found near the origin of the spectra in Fig.\ \ref{Fig2} and Figs.\ \ref{Fig8}-\ref{Fig10}.

\subsection{Case I: monthly eclipses}
The first case is for $\alpha r<s_1-s_2$,
where there is a complete lunar eclipse and a complete planetary eclipse every month: at $t=t_k$ the moon becomes fully covered by planet shadow and at $t=\bar t_k$ the moon shadow falls completely onto the planet. The planetary eclipses can happen when the planet is bright and the moon is dark. The corresponding eclipse magnitude $\bar\Upsilon$ is given by (\ref{defM}) with $z_-=l-s_2$ and $z_+=l+s_2$. With the substitution $l=l_k$ from (\ref{ln}), this gives
\[
\bar\Upsilon(\alpha r\sin\vartheta) = 3\pi {s_2}^2 \frac{4s_1^2-s_2^2-4\alpha^2 r^2\sin^2\vartheta}{16vs_1^3}
.
\]
There are three nonzero coefficients:
\[
\bar\Upsilon_0 = 3\pi {s_2^2} \frac{4s_1^2-s_2^2-2\alpha^2 r^2}{16vs_1^3}
, \quad
\bar\Upsilon_{\pm 2} = -3\pi {s_2}^2 \frac{\alpha^2 r^2}{16vs_1^3}
.
\]
We now use $\bar g_n=-\sum_m\bar\Upsilon_m {f_{\mathrm D}}_{n-m}$ and find peak amplitudes:
\[
{f_\mathrm{E}}^0_n = -\frac{v{s_1}^2}{8\pi R^2 r}(h_n \Upsilon_0 + h_{n-2} \Upsilon_2 + h_{n+2} \Upsilon_2)
.
\]
If there is no inclination, then $\alpha=0$ and $\bar\Upsilon_2=0$ and the side bands are small copies of the direct spectrum found at $m=0$. This case was plotted in Fig.\ \ref{Fig3}. The situation when the moon is bright and the planet is dark is very similar, because the eclipse magnitude function has a similar shape, only slightly larger in the center (see Fig.\ \ref{Fig5}), however the equations are not as simple.

\subsection{Cases II and III: partial eclipses}
When there are partial eclipses, these always occur in a specific range of orbital phases. For cases II and III, the eclipses are incomplete in the intervals $\vartheta_1<|\vartheta|<\pi-\vartheta_1$, modulo $2\pi$. 
Here, $\vartheta_1$ is the (smallest) angle where the overlapping disks are touching at the poles: here, $\alpha r\sin\vartheta_1=s_1-s_2$. There are four of these contact points along the orbit of the barycenter. Because the magnitude functions $\Upsilon$ and $\bar\Upsilon$ are not smooth here, these points determine the behavior of the Fourier tails of ${f_\mathrm{E}}^m_n$ for cases II and III. In Appendix \ref{AppB} we derive these tails. For case III, new (sub)intervals $\vartheta_2<|\vartheta|<\pi-\vartheta_2$ modulo $2\pi$ appear, with $\vartheta_2$ being the smallest solution of $\alpha r\sin\vartheta_2=s_1+s_2$. At the boundaries, four extra contact points along the orbit are found. Here the disks have no overlap and 
touch at the poles. These also turn up in the Fourier tails.

\begin{figure}
\centering
\resizebox{\hsize}{!}{
\includegraphics[width=9cm]{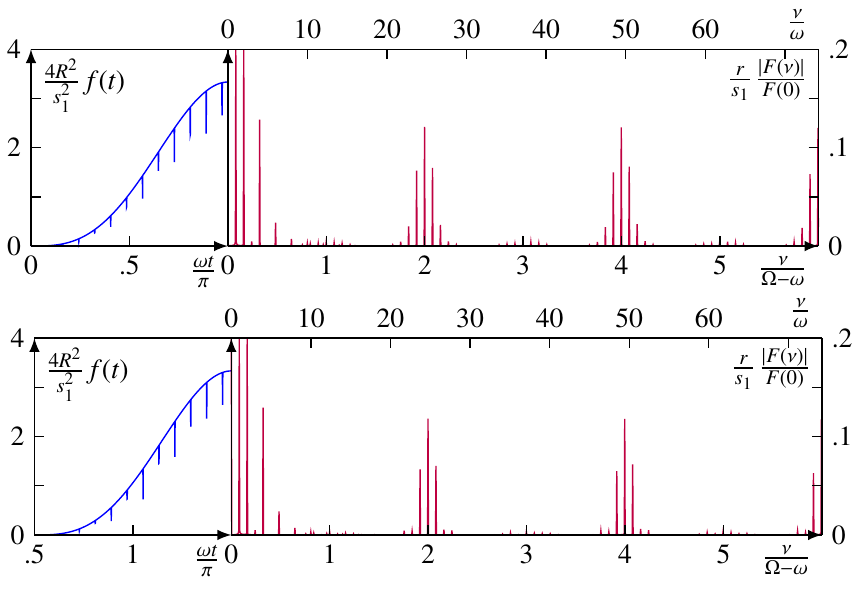}
}
\caption{
\label{Fig8}
Signals and spectra for the eclipses in an inclined system. The inclination angle is $\alpha=(s_1-s_2)/r=s_1/2r$. This is the maximum value where complete eclipses always occur (case I). Same planet radii and axes as in Fig.\ \ref{Fig3}. 
Top: Observation direction is $\hat{\boldsymbol{o}}=\boldsymbol{i}$, along the line of nodes.
Bottom: Observation in direction $\hat{\boldsymbol{o}}=\boldsymbol{j}$. The eclipses are on the equator for $\omega t_k\approx\pi$. This is at the full phase and quarter phase, for top and bottom figures, respectively.
Because there is an complete eclipse twice a month, there is little difference in the signals (see Fig.\ \ref{Fig5}).
}
\end{figure}

\begin{figure}
\centering
\resizebox{\hsize}{!}{
\includegraphics[width=9cm]{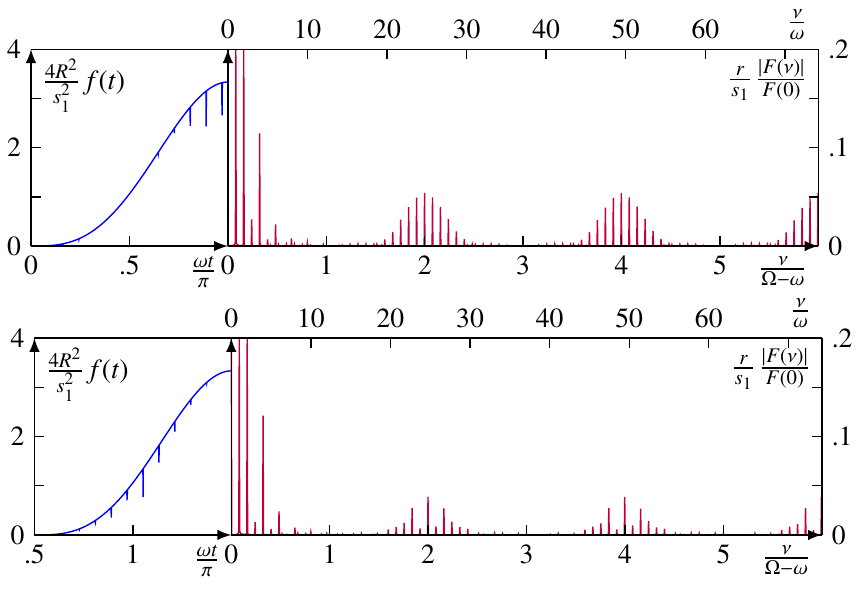}
}
\caption{
\label{Fig9}
Signals and spectra for $\alpha=(s_1+s_2)/r=3s_1/2r$. This is the maximum value where (partial) eclipses always occur (case II); they appear strongest around $\omega t_k=\pi$.
The side bands in the spectrum are broadened due to the diminishing strength of the eclipses around other times. Top and bottom: $\hat{\boldsymbol{o}}=\boldsymbol{i}$ and $\hat{\boldsymbol{o}}=\boldsymbol{j}$, with the same axes as in Fig.\ \ref{Fig3}. In the bottom figure the strongest eclipses (observed at quarter phase) are weaker compared to those (observed at full phase) in the top figure.
}
\end{figure}

\begin{figure}
\centering
\resizebox{\hsize}{!}{
\includegraphics[width=9cm]{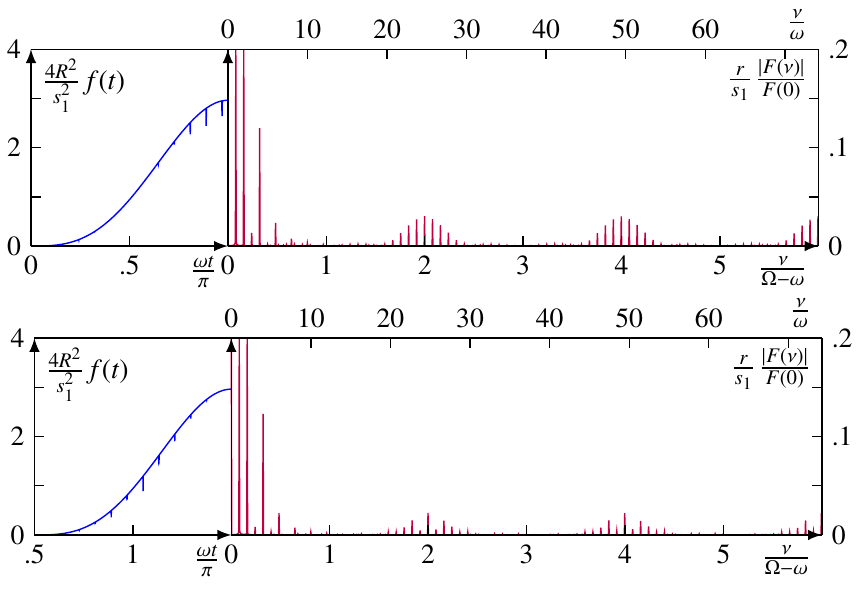}
}
\caption{
Signals and spectra for $s_1/s_2=3$ and $\alpha=(s_1+s_2)/r=4s_1/3r$. This is again the maximum value where partial eclipses always occur. All other parameters are kept the same as Fig.\ \ref{Fig8}. The eclipse contributions to the signals are weaker by roughly a factor of 
$4/9$
compared to Fig.\ \ref{Fig9}.
}
\label{Fig10}
\end{figure}

Figures \ref{Fig8} and\ \ref{Fig9} show the Fourier spectra for edge-on observation for two special values of inclination; these respectively show full eclipses occurring every month and partial eclipses every month. For observation along the nodes ($\hat{\boldsymbol o}=\boldsymbol i$), the eclipse signal is strongest. The case where there is not always an eclipse
is not plotted: then the side bands flatten out even more. Figure\ \ref{Fig10} shows the same situation as Fig.\ \ref{Fig9}, but for different planet radii. 

\begin{table}
\caption{Estimates for orders of magnitude of Fourier-peak strengths.}
\label{table2}      
\centering
\begin{tabular}{l|l|lll}
\hline
\hline
system & & & HZ Jupiter & HZ Earth \\
& & Jupiter & M-dwarf & Sun-type \\
orbital axis & $R$ & $10S$ & $10^2S$ & $10^3S$  \\
lunar axis & $r$ & $S$ & $10S$ & $S$  \\
\hline
\hline
effect & $|{f_1}^0|$ &&&  \\
\hline
transit & $\displaystyle\frac{s^2}{SR}$ & $10^{-1}$ & $10^{-2}$ & \textcolor{blue}{$10^{-7}$} \\
reflection & $\displaystyle\frac{s^2}{R^2}$ & $10^{-2}$ & \textcolor{blue}{$10^{-4}$} & \textcolor{blue}{$10^{-10}$} \\
binary transit & $\displaystyle\frac{s_1{s_2}^2}{rR^2}$ & $10^{-2}$ & \textcolor{blue}{$10^{-5}$} & \textcolor{blue}{$10^{-12}$} \\
binary eclipse & $\displaystyle\frac{s_1{s_2}^2}{rR^2}$ & $10^{-2}$ & \textcolor{blue}{$10^{-5}$} & \textcolor{blue}{$10^{-12}$} \\
second reflection & $\displaystyle\frac{{s_1}^2{s_2}^2}{r^2R^2}$ & $10^{-2}$ & \textcolor{blue}{$10^{-6}$} & \textcolor{red}{$10^{-14}$} \\
lunar tide & $\displaystyle\frac{{s_1}^3{s_2}^2}{r^3R^2}$ & $10^{-2}$ & \textcolor{blue}{$10^{-7}$} & \textcolor{red}{$10^{-16}$} \\
stellar tide & $\displaystyle\frac{s^3}{R^3}$ & $10^{-3}$ & \textcolor{red}{$10^{-6}$} & \textcolor{red}{$10^{-15}$} \\
planet tide & $\displaystyle\frac{s^2S^3}{R^5}$ & \textcolor{blue}{$10^{-4}$} & \textcolor{red}{$10^{-9}$} & \textcolor{red}{$10^{-18}$} \\
\hline
\hline
\end{tabular}
\tablefoot{Red and blue numbers indicate swamping by stellar noise at a level of $10^{-3}$, with and without occultor, respectively. We assumed an observation duration $T$ of one orbital period; longer observation increases the values of $|{f_n}^m|$. Bottom three rows estimate strengths from the gravitational tide of a nearby moon on the planet, of the planet on the star, and of the star on the planet respectively.}
\end{table}

\section{Double reflections}
\label{SecVI}
When both the planet and its moon have high albedo, light rays could bounce off one planet and then off its companion, before being reflected towards the observer. These secondary reflections also contribute to the monthly modulation in the light curve. We now show that the effect of double reflection in the Fourier spectrum could become comparable to eclipses at higher inclinations $\alpha$ when eclipses are rare.

Consider the stellar light that is first reflected off the planet then off the moon. If we take unit albedos $a_1=a_2=1$, the lowest-order contribution in $s_1/r$ to the net signal is:
\[
\frac{s_1^2}{\pi R^2}
\left( \int\int_{\rightmoon} \mathrm{d}^2\hat{\boldsymbol s} (\hat{\boldsymbol R}\bullet\hat{\boldsymbol s}) (\hat{\boldsymbol s}\bullet\hat{\boldsymbol r}) \right)
\frac{s_2^2}{\pi r^2}
\left( \int\int_{\rightmoon} \mathrm{d}^2\hat{\boldsymbol s} (\hat{\boldsymbol r}\bullet\hat{\boldsymbol s}) (\hat{\boldsymbol s}\bullet\hat{\boldsymbol o}) \right)
.
\]
The spherical lunes on the unit sphere are the same for both planet and moon. We find a factor of 
$4/3$ for the $z$ integrals, and obtain
\[
\bigg(\frac{4s_1s_2}{3\pi Rr}\bigg)^2
\int\limits_0^{2\pi} \mathrm{d}\theta\, c(\theta-\vartheta-\pi)c(\theta-\varphi)
\int\limits_0^{2\pi} \mathrm{d}\theta\, c(\theta-\varphi-\pi) c(\theta-\phi_\mathrm{o})
.
\]
We now express both these integrals in terms of the single-planet phase curve $h$. When we add the contribution for light that is first scattered off the moon and then off the plane, we obtain the net double-reflection
\[
f_\mathrm{S} = \bigg(\frac{s_1s_2}{4Rr}\bigg)^2 \Big[ h(\vartheta-\varphi)h(\varphi-\phi_\mathrm{o}) + h(\vartheta-\varphi-\pi)h(\varphi-\phi_\mathrm{o}+\pi) \Big]
.
\]
For the Fourier coefficients, we can now use 
\[
\frac{1}{(2\pi)^2} \int\limits_0^{2\pi}\mathrm{d}\vartheta \int\limits_0^{2\pi}\mathrm{d}\varphi\, \mathrm e^{-\mathrm in\vartheta-\mathrm im\varphi} h(\vartheta-\varphi)h(\varphi-\phi_\mathrm{o}) =
\mathrm e^{-\mathrm i(n+m)\phi_\mathrm{o}} h_n h_{n+m}
.
\]
Hence
\begin{equation}
{f_\mathrm{S}}^m_n =  \bigg(\frac{s_1s_2}{4Rr}\bigg)^2 2\mathrm e^{-\mathrm i(n+m)\phi_\mathrm{o}} h_n h_{n+m}
, \quad m\ \mathrm{even}
,
\label{fM}
\end{equation}
and ${f_\mathrm{S}}^m_n=0$ if $m$ is odd. 
If one compares the first two side bands at $m=2$ and $m=4$, this pattern is characterized by ${f_0}^4$, ${f_2}^4$, ${f_4}^4$, being a copy of ${f_2}^2$, ${f_4}^2$, ${f_6}^2$ but smaller. Similarly, the harmonic $m=6$ from the monthly revolutions have the same relative sizes but again smaller. The amplitudes are all positive for $\phi_\mathrm{o}=0$.
The values are pretty low: even for unit albedo the dominant side peak $f^2_{-1}$ is a factor $3(\pi s_1/12r)^2$ weaker than the main peak in the direct reflection.

\begin{table}
\caption{Asymptotic power-law behavior of the Fourier tails.}
\label{table3}
\centering
\begin{tabular}{l|ll}
\hline
\hline
effect & Lambertian & ocean glint \\
\hline
direct reflection & $n^{-4}$ & $n^{-2}$ \\
full eclipses & $-(n+m)^{-4}$ & $-(n+m)^{-2}$ \\
partial eclipses & $-(n+m)^{-7/2}$ & $-(n+m)^{-3/2}$ \\
double reflection & $n^{-4}(n+m)^{-4}$ & $n^{-2}(n+m)^{-2}$ \\
\hline
\hline
\end{tabular}
\tablefoot{
The first column is for a homogeneous Lambertian system; the second is for the glint from a circumventing ocean \citep{Visser2015}. 
The magnitude for the eclipsed glint is proportional to the cord length of the shadow disk (of the planet) on the equator (of the moon). Since this length behaves as a square root, $k=3/2$ (see Appendix\ \ref{AppB}). These results are for edge-on observation; the order of magnitude is suppressed.}
\end{table}

\section{Conclusions}
\label{SecVII}
Because planets have very well-defined orbital periods, the Fourier transform of the phase light curve of an exoplanet system will consist of sharp peaks (but broadened by the intensity spectrum of the host star). Each planet contributes an equidistant set and the individual sets do not fall on top of each other because the periods of different planets are generally incommensurable. This allows the astronomer to separate contributions from different planets.

Here, we study the reflection signal from a planet with a moon. The signal is double-periodic of the form (\ref{fourier}) with the basis frequencies $\omega$ and $\Omega$.
These frequencies now correspond to the annual (barycenter) motion and the lunar (relative) motion.
We consider homogeneous companions, with unit (or zero) albedo for near edge-on observation and a small inclination $\alpha$ of the lunar plane.
We show that if the radii $s_1$, $s_2$ are small compared to the planet separation $r$, the Fourier series has a unique form. According to  (\ref{fDresult}), (\ref{fEresult}), (\ref{hIdef}) and (\ref{fM}), now including albedo-factors $a_1$ and $a_2$, this form is:
\begin{align*}
& f_\mathrm{D}(t) = \frac{s_1^2 a_1+s_2^2 a_2}{4R^2}\sum_n \mathrm e^{\mathrm in\omega t} \mathrm e^{-\mathrm in\phi_\mathrm{o}} h_n
, \\
& f_\mathrm{E}(t) = 
{\frac{v}{2\pi r}} \sum_{nm} \mathrm e^{\mathrm i(n\omega + m\Omega) t} \mathrm e^{-\mathrm i(n+m)\phi_\mathrm{o}} \Big[ a_1 g_{n+m} + (-1)^m a_2 \bar g_{n+m} \Big]
, \\
& f_\mathrm{S}(t) = \bigg(\frac{s_1s_2}{4Rr}\bigg)^2 a_1a_2 \sum_{nm} \mathrm e^{\mathrm i(n\omega + m\Omega) t} \mathrm e^{-\mathrm i(n+m)\phi_\mathrm{o}} \Big[ 1 + (-1)^m \Big] h_n h_{n+m}
.
\end{align*}
The coefficients $h_n$ are for the phase light curve $h$ of a single homogeneous Lambertian planet, given by (\ref{hn}).

Because two companions have the same orbital phase $\vartheta$, the direct reflection $f_\mathrm{D}$ cannot reveal a planet binary. The peaks that show the binary are found at multiples of $\Omega$ in the weaker terms $f_\mathrm{E}$ and $f_\mathrm{S}$. These Fourier series have typical features. Small but repeated identical side bands arise from the eclipses on the planet binary. They do not diminish in strength for $|m|\ll r/s_1$. The coefficients ${f_\mathrm{E}}^0_n$ for eclipses are found from (\ref{pattern}) in terms of the function $g$ defined in (\ref{delta}).
Partial eclipses give the characteristic power-law $n^{-7/2}$ tail, coming from the contact points. If many peaks can be resolved, this asymptotic behavior may be useful. The double reflection between companions also gives side bands that are identical in shape, but instead decay in strength with the $m$-value.

When the two planets are so close that $s_1\lesssim r$, our analytic approach breaks down. The three effects obtain comparable signal strength and the decomposition (\ref{decomp}) becomes problematic. It is no longer possible to attribute the $m$ side bands to one effect.

\bibliographystyle{aa}
\bibliography{references}{}

\begin{thebibliography}{31}
\expandafter\ifx\csname natexlab\endcsname\relax\def\natexlab#1{#1}\fi

\bibitem[{Barr(2016)}]{Barr2016}
Barr, A. 2016, The Astronomical Review, 12, 24

\bibitem[{Belikov {et~al.}(2015)Belikov, Bendek, Thomas, Males, \&
  Lozi}]{Belikov2015}
Belikov, R., Bendek, E., Thomas, S., Males, J., \& Lozi, J. 2015, in Proc.SPIE,
  Vol. 9605, Techniques and Instrumentation for Detection of Exoplanets VII,
  960517

\bibitem[{Boccaletti {et~al.}(2004)Boccaletti, Baudoz, Baudrand, Reess, \&
  Rouan}]{Boccaletti2004}
Boccaletti, A., Baudoz, P., Baudrand, J., Reess, J., \& Rouan, D. 2004,
  Advances in Space Research, 36, 1099–1106

\bibitem[{Borucki {et~al.}(2009)Borucki, Koch, Jenkins, Sasselov, Gilliland,
  Batalha, Latham, Caldwell, Basri, Brown, Christensen-Dalsgaard, Cochran,
  DeVore, Dunham, Dupree, Gautier, Geary, Gould, Howell, Kjeldsen, Lissauer,
  Marcy, Meibom, Morrison, \& Tarter}]{Borucki2009}
Borucki, W.~J., Koch, D., Jenkins, J., {et~al.} 2009, Science, 325, 709

\bibitem[{Cabrera \& Schneider(2007{\natexlab{a}})}]{CabreraSchneider2007}
Cabrera, J. \& Schneider, J. 2007{\natexlab{a}}, \aap, 464, 1133

\bibitem[{Cabrera \& Schneider(2007{\natexlab{b}})}]{Cabrera2007}
Cabrera, J. \& Schneider, J. 2007{\natexlab{b}}, in Astronomical Society of the
  Pacific Conference Series, Vol. 366, Transiting Extrapolar Planets Workshop,
  ed. C.~Afonso, D.~Weldrake, \& T.~Henning, 242

\bibitem[{Cash(2006)}]{Cash2006}
Cash, W. 2006, \nat, 442, 51

\bibitem[{Cowan \& Agol(2008)}]{Cowan2008}
Cowan, N. \& Agol, E. 2008, \apj, 678, L129

\bibitem[{Cowan {et~al.}(2013)Cowan, Fuentes, \& Haggard}]{Cowan2013}
Cowan, N., Fuentes, P., \& Haggard, H. 2013, \mnras, 434, 2465

\bibitem[{Douglas {et~al.}(2018)Douglas, Carlton, Cahoy, Jeremy~Kasdin,
  Turnbull, \& B.}]{Douglas2018}
Douglas, E., Carlton, A., Cahoy, K.~L., {et~al.} 2018, in Proc.SPIE, Vol.
  10705, 1070526--1

\bibitem[{Fujii \& Kawahara(2012)}]{Fujii2012}
Fujii, Y. \& Kawahara, H. 2012, \apj, 755, 101

\bibitem[{Gillon {et~al.}(2017)Gillon, Triaud, Demory, Jehin, Agol, Deck,
  Lederer, de~Wit, Burdanov, Ingalls, Bolmont, Leconte, Raymond, Selsis,
  Turbet, Barkaoui, Burgasser, Burleigh, Carey, Chaushev, Copperwheat, Delrez,
  Fernandes, Holdsworth, Kotze, Van~Grootel, Almleaky, Benkhaldoun, Magain, \&
  Queloz}]{Gillon2017}
Gillon, M., Triaud, A., Demory, B.-O., {et~al.} 2017, \nat, 542, 456

\bibitem[{Heller {et~al.}(2016)Heller, Michael~Hippke, Ben~Placek, Angerhausen,
  \& Eric~Agol}]{HellerHippke2016}
Heller, R., Michael~Hippke, M., Ben~Placek, B., Angerhausen, D., \& Eric~Agol,
  E. 2016, \aap, 591, A67

\bibitem[{Kane \& Gelino(2013)}]{Kane2013}
Kane, S.~R. \& Gelino, D.~M. 2013, \apj, 762, 129

\bibitem[{Kipping(2011)}]{Kipping2009}
Kipping, D. 2011, \mnras, 416, 689

\bibitem[{Krist {et~al.}(2007)Krist, Beichman, Trauger, Rieke, Somerstein,
  Green, Horner, Stansberry, Shi, Meyer, Stapelfeldt, \& Roellig}]{Krist2007}
Krist, J., Beichman, C., Trauger, J., {et~al.} 2007, Proc. SPIE, 6693, 6693OH1

\bibitem[{Lewis {et~al.}(2015)Lewis, Ochiai, Nagasawa, \& Ida}]{Lewis2015}
Lewis, K., Ochiai, H., Nagasawa, M., \& Ida, S. 2015, \apj, 805, 27

\bibitem[{Link(1969)}]{Link1969}
Link, F. 1969, Eclipse Phenomena in Astronomy, 1st edn. (Berlin Heidelberg:
  Springer-Verlag)

\bibitem[{Lovis {et~al.}(2011)Lovis, S{\'e}gransan, Mayor, Udry, Benz, Bertaux,
  Bouchy, Correia, Laskar, Lo~Curto, Mordasini, Pepe, Queloz, \&
  Santos}]{Lovis2011}
Lovis, C., S{\'e}gransan, D., Mayor, M., {et~al.} 2011, \aap, A112

\bibitem[{Mawet {et~al.}(2010)Mawet, Serabyn, Liewer, Burruss, Hickey, \&
  Shemo}]{Mawet2010}
Mawet, D., Serabyn, E., Liewer, K., {et~al.} 2010, \apj, 709, 53–57

\bibitem[{Morse {et~al.}(2018)Morse, Bendek, Cabrol, Marchis, Turnbull,
  Chakrabarti, Fischer, Goldblatt, Guyon, Hart, Males, \& Kasting}]{Morse2018}
Morse, J., Bendek, E., Cabrol, N., {et~al.} 2018, ArXiv e-prints
  [\eprint[arXiv]{1803.04872}]

\bibitem[{Namouni(2010)}]{Namouni2010}
Namouni, F. 2010, \apjl, 719, L145

\bibitem[{Ochiai {et~al.}(2014)Ochiai, Nagasawa, \& Ida}]{Ochiai2014}
Ochiai, H., Nagasawa, M., \& Ida, S. 2014, \apj, 790, 92

\bibitem[{Ogihara \& Ida(2012)}]{Ogihara2012}
Ogihara, M. \& Ida, S. 2012, \aj, 753, 60

\bibitem[{Olver {et~al.}(2010)Olver, Lozier, Boisvert, \& Clark}]{Olver2010}
Olver, F., Lozier, D., Boisvert, R., \& Clark, C. 2010, NIST Handbook of
  Mathematical Functions (Cambridge: Cambridge University Press)

\bibitem[{Pannekoek(1947)}]{Pannekoek1947}
Pannekoek, A. 1947, Popular Astronomy, 55, 422

\bibitem[{Shallue \& Vanderburg(2018)}]{Shallue2018}
Shallue, C. \& Vanderburg, A. 2018, \aj, 155, 94

\bibitem[{Snellen {et~al.}(2009)Snellen, de~Mooij, \& Albrecht}]{Snellen2009}
Snellen, I., de~Mooij, E., \& Albrecht, S. 2009, \nat, 459, 543

\bibitem[{van Hulst(1980)}]{vanHulst1980}
van Hulst, H. 1980, Multiple Light Scattering: Tables, Formulas, and
  Applications, Vols. 1 and 2 (New York: Academic Press)

\bibitem[{Vep\v{s}tas(2008)}]{Vepstas2008}
Vep\v{s}tas, L. 2008, Numerical Algorithms, 47, 211–252

\bibitem[{Visser \& van~de Bult(2015)}]{Visser2015}
Visser, P. \& van~de Bult, F. 2015, \aap, 579

\end{thebibliography}

\appendix
\section{Condition for absence of transits}
\label{AppA}
A transit happens if the starlight in the direction of the observer is blocked by a planet. The disk of the planet overlaps the disk of the host star in the (projected) plane of observation.
Defining $l$ to be the distance between the centers of the two disks, transits of planet 1 occur for
\[
l = |\hat{\boldsymbol o}\times \boldsymbol R_1| < S \pm s_1 , \quad \hat{\boldsymbol o}\bullet\boldsymbol R_1 > 0
.
\]
The plus sign is for the partial transit (planet disk is partially in front of the star disk) and the minus sign is for a complete transit (planet disk is inside the star disk). The component of $\boldsymbol R_1$ in the direction is positive when the planet has to be in between the observer and the star. An occultation of the planet by the star occurs for a negative dot product. If we assume that $\omega$ and $\Omega$ are incommensurable, the value of $l$ is minimized for
\[
\hat{\boldsymbol o}\bullet\frac{\mathrm d \boldsymbol R}{\mathrm d t} = 0 , \quad \hat{\boldsymbol o}\bullet\frac{\mathrm d \boldsymbol r}{\mathrm d t} = 0
.
\]
These equations imply $\omega t=\phi_\mathrm{o}+n\pi$ (see Fig.\ \ref{Fig1}).
The minimal value of the displacement is $l=|\hat{\boldsymbol o}\bullet\boldsymbol k|R + |\hat{\boldsymbol o}\bullet\hat{\boldsymbol e}_3|r_1 =$
\[
R\cos\theta_\mathrm{o} + r_1(\cos\alpha\cos\theta_\mathrm{o}-\sin\alpha\sin\theta_\mathrm{o}\sin\phi_\mathrm{o})
,
\]
with $r_1=|\boldsymbol R_1-\boldsymbol R|=m_2 r/(m_1+m_2)$ and $r_2=r-r_1$. The condition for transits never to occur is, for $\theta_\mathrm{o}\lessapprox\frac{\pi}{2}$ and $0\lessapprox\alpha$:
\[
l = (R + r_1)(\tfrac{\pi}{2} - \theta_\mathrm{o}) - r_1\alpha \sin\phi_\mathrm{o} > S + s_1
.
\]
Because we are considering two planets, we have the two conditions for the observer inclination:
\[
\theta_\mathrm{o} < \frac{\pi}{2} - \frac{S+s_1+r_1\alpha\sin\phi_\mathrm{o}}{R-r_1} , \quad 
\theta_\mathrm{o} < \frac{\pi}{2} - \frac{S+s_2+r_2\alpha\sin\phi_\mathrm{o}}{R-r_2}
.
\]

In this paper, we also assume that due to inclined observation the planet and moon also never block the direct reflected light towards the observer. We now derive the required condition for absence of these types of mutual events. Let $l$ be the distance between the disk-centers of planet and moon projected onto the plane of observation (i.e.,\ the celestial plane). It then follows that $l$ is the length of the component of the distance vector $\boldsymbol r$ orthogonal to $\hat{\boldsymbol o}$. This is $l=|\boldsymbol r-\hat{\boldsymbol o}(\hat{\boldsymbol o}\bullet\boldsymbol r)|$. Hence, the planet disk is in front of the moon disk for
\[
l = |\hat{\boldsymbol o}\times \boldsymbol r| < s_1 + s_2 , \quad \hat{\boldsymbol o}\bullet\boldsymbol r < 0
,
\]
and the moon disk is in front of the planet disk for
\[
l = |\hat{\boldsymbol o}\times \boldsymbol r| < s_1 + s_2 , \quad \hat{\boldsymbol o}\bullet\boldsymbol r > 0
.
\]
With the plus sign in these expressions replaced by a minus sign, one obtains the condition for the larger disk completely overlapping the smaller disk. The lowest value of $l$ occurs for a difference velocity perpendicular to the observation direction, or for
\[
\hat{\boldsymbol o}\bullet\frac{\mathrm d \boldsymbol r}{\mathrm d t} = 0
.
\]
This implies, using our assumption (\ref{circles}) that
\[
\sin(\Omega t - \phi_\mathrm{o}) = 
(\cos\alpha-1)\cos\Omega t\sin\phi_\mathrm{o}
,
\]
or
\[
\Omega t = \phi_\mathrm{o} + \arctan\frac{(\cos\alpha-1)\sin\phi_\mathrm{o}\cos\phi_\mathrm{o}}{\cos^2\phi_\mathrm{o}+\cos\alpha\sin^2\phi_\mathrm{o}}
 + n\pi
.
\]
For these phases, 
the minimal displacement is
\[
l = |\hat{\boldsymbol o}\bullet\hat{\boldsymbol e}_3|r
.
\]
The transits do not take place if we demand $l>s_1+s_2$. For small inclinations $\alpha$, the condition for observer inclination with respect to the orbital plane becomes
\[
\theta_\mathrm{o} < \frac{\pi}{2} - \frac{s_1 + s_2}{r} - \alpha \sin\phi_\mathrm{o}
.
\]

\section{Fourier tails for partial eclipses}
\label{AppB}
It is well known that the Fourier coefficients $f_n$ of a periodic function $f$ that is also an analytic function decay with $n$ faster than any power law. Therefore, the behavior around the points where a function is not analytic determines the asymptotic behavior of the Fourier coefficients. The direct light curve $f_\mathrm{D}$ of a planet is only not analytic for $\vartheta=\phi_0$ when the planet is at inferior conjunction. Similarly, the periodic function $g$ (describing $f_\mathrm{E}$) is not analytic at $l=\pm s_1\pm s_2$. This is when an edge of the shadow band just touches its companion (at a pole). We here calculate the Fourier tail for $f_\mathrm{D}$ and $f_\mathrm{E}$.

In order to study the behavior as $x\longrightarrow 0^+$ of a noninteger power $x^{k-1}$ with $k>1$, like $\sqrt{x}$,
we need the Hurwitz zeta function $\zeta(s,x)$. By isolating the branch point at $x=0$, the Hurwitz function can be written with the series \citep{Vepstas2008}:
\[
\zeta(1-k,x) =  x^{k-1} + \sum_{n=0}^{\infty} \binom{n-k}{n}\zeta(1+n-k)(-x)^n
.
\]
The analytic part of the Hurwitz function on $[0,1]$ has series coefficients determined by the Riemann zeta function $\zeta(s)$. This part is repeated at $x=1$ 
\[
\zeta(1-k,x) =  \sum_{n=0}^{\infty} \binom{n-k}{n}\zeta(1+n-k)(-x+1)^n
.
\]
For arguments on the $x$-interval $[0,1]$ and $k>1$, the Fourier series of the above function is \citep{Olver2010}
\[
\zeta(1-k,x) = \frac{\Gamma(k)}{(2\pi)^k} \bigg( \mathrm e^{\mathrm i\pi k/2} \sum_{n=1}^\infty\frac{\mathrm e^{-2\pi\mathrm inx}}{n^k} + \mathrm e^{-\mathrm i\pi k/2} \sum_{n=1}^\infty\frac{\mathrm e^{2\pi\mathrm inx}}{n^k}\bigg)
.
\]
This is Hurwitz's formula. We note that the branch point determines the asymptotic behavior of the Fourier coefficients. For integer $k$, the Taylor series terminates and becomes the Bernoulli polynomial: $\zeta(1-k,x)=-B_k(x)/k$. Although the coefficient for $x^{k-1}$ is now $1-\tfrac{1}{2}$, its $(k-1)$-th derivative still has a step-discontinuity of size $(k-1)!$. The periodic continuation is then also given by the above Fourier series.

We now consider
a periodic function $f(\theta)$, with period $2\pi$ and with nonanalytic behavior around one point, as in
\[
f(\theta) = \left\{ \begin{array}{ll} 
(\theta-a)^{k-1}A + \cdots , & \quad \theta \longrightarrow a^+ , \\
(a-\theta)^{l-1}B + \cdots , & \quad \theta \longrightarrow a^- ,
\end{array} \right.
\]
with $k>1$, $l>1$. By comparing with the Hurwitz function, we find that $f(\theta)$ has Fourier coefficients with tails
\begin{equation}
f_n = \frac{\mathrm e^{-\mathrm ina}}{2\pi} \times \left\{ \begin{array}{ll} \displaystyle
A\Gamma(k)\frac{\mathrm e^{-\mathrm i\pi k/2}}{n^k} + B\Gamma(l)\frac{\mathrm e^{\mathrm i\pi l/2}}{n^l} + \cdots , & \quad n \longrightarrow\infty ,
\\ \displaystyle
A\Gamma(k)\frac{\mathrm e^{\mathrm i\pi k/2}}{(-n)^k} + B\Gamma(l)\frac{\mathrm e^{-\mathrm i\pi l/2}}{(-n)^l} + \cdots , & \quad n \longrightarrow -\infty .
\end{array} \right.
\label{Hurwitz}
\end{equation}
For integer $k=l$ and $B=(-1)^{k-1}A$ the expression vanishes, because then the function is analytic and (\ref{Hurwitz}) does not apply. For integer $k=l\geq 2$ and $B=(-1)^kA$, the correspondence becomes
\begin{equation}
f(\theta) = (\theta-a)^{k-2}|\theta-a|A + \cdots
, \quad
f_n = 2A \frac{(k-1)!\mathrm e^{-\mathrm ina}}{2\pi(\mathrm in)^k} + \cdots
.
\label{intl}
\end{equation}
We illustrate the method with the direct phase curve (\ref{fDresult}). It is everywhere three times differentiable, except at $\theta=0$. Near $\theta=0$, it behaves as $h(\theta)\approx 8|\theta|^3/9\pi$, so that its third derivative jumps. Equation (\ref{intl}) then provides the asymptote $h_n\approx 16/3\pi^2 n^4$. The value given by (\ref{hn}) is twice as large. However $h(\theta)-4(\cos\theta)/3$ actually has periodicity $\pi$, not $2\pi$. By taking into account that almost all the coefficients are even, one finds the correct asymptote.

For the eclipses, we shall need the asymptotes for the periodic function $\Upsilon(\alpha r\sin\vartheta)$. We consider the case II of planets that always eclipse but the eclipses can be partial: $s_1-s_2<\alpha r<s_1+s_2$. The function $\Upsilon$ is everywhere three times differentiable, except at $l=\pm (s_1-s_2)$ where the third derivative becomes infinite. This occurs when the shadow of the planet touches a pole of the moon at the eclipse maximum, which happens for four values of $\vartheta$. Let the orbital phase $\vartheta_1$ be defined by one solution of $\alpha r\sin\vartheta_1=s_1-s_2$.
According to (\ref{fEresult}), the peaks are determined by the Fourier transform of $g(\vartheta)=-f_\mathrm{D}(\vartheta/\omega)\Upsilon(\alpha r\sin\vartheta)$. This function is nonanalytic where either of the functions $f_\mathrm{D}$ or $\Upsilon$ is not analytic. Although $f_\mathrm{D}$ is nonanalytic with $k=4$, the effect in $\Upsilon$ has power $k=7/2$ with is dominant (for large $n$). We therefore approximate
\begin{equation}
{f_\mathrm{E}}^0_n = \frac{-v}{2\pi r} \Big( f_\mathrm{D}(\tfrac{\vartheta_1}{\omega}) + f_\mathrm{D}(\tfrac{-\vartheta_1}{\omega}) + 
f_\mathrm{D}(\pi+\tfrac{\vartheta_1}{\omega}) + f_\mathrm{D}(\pi-\tfrac{\vartheta_1}{\omega})
\Big) \Upsilon_n
,
\label{gn}
\end{equation}
with $\Upsilon_n$ the Fourier coefficients of $\Upsilon$ considered as periodic function of $\vartheta$. For the lunar eclipses, the magnitude is given by (\ref{defM}). Approaching the contact point $l=s_2-s_1$ from above, it behaves as
\[
\Upsilon(l) = \frac{4\!\sqrt{2s_1}}{5vs_2^2} (l-s_2+s_1)^{5/2} + \cdots , \quad l\longrightarrow (s_2-s_1)^+
.
\]
The dots now also contain the analytic part. The function has no fractional powers in the expansion for $l\longrightarrow (s_2-s_1)^-$. Of course, $\Upsilon$ is an even function of $l$. Substituting $l=\alpha r\sin\vartheta$, we now evaluate the coefficients in the tail:
\begin{align}
{f_\mathrm{E}}^0_n = 
& \frac{-3\!\sqrt{2s_1}}{2\pi^{3/2} rs_2^2} \bigg(\frac{s_1-s_2}{\tan\vartheta_1}\bigg)^{5/2} \frac{\cos(|n|\vartheta_1+\tfrac{\pi}{4})}{|n|^{7/2}} \times
\nonumber \\
& \Big( f_\mathrm{D}(\tfrac{\vartheta_1}{\omega}) + f_\mathrm{D}(\tfrac{-\vartheta_1}{\omega}) + 
f_\mathrm{D}(\pi+\tfrac{\vartheta_1}{\omega}) + f_\mathrm{D}(\pi-\tfrac{\vartheta_1}{\omega}) \Big)
,
\label{expand}
\end{align}
and $n$ even. The
magnitude $\bar\Upsilon$ for planetary eclipses also jumps at $l=\pm(s_1-s_2)$. The jump discontinuity has the same expression, except with $s_1$ and $s_2$ interchanged. An interesting case occurs for planets of equal size and albedo. If $s_2\longrightarrow s_1$, then $\vartheta_1\longrightarrow 0$. For unit albedo, the combined spectrum for both planets eclipsing each other every halve month becomes, for both $n$ and $m$ even:
\[
{f_\mathrm{E}}^m_n = -\frac{3\alpha^{5/2}\!\sqrt{r^3s_1}}{8\pi^{3/2}R^2|n+m|^{7/2}} \Big( h(\phi_\mathrm{o})+ h(\pi-\phi_\mathrm{o}) \Big)
.
\]

In the final case III without monthly eclipses:
$s_1+s_2<\alpha r$. The orbital phase $\vartheta_2$
above which eclipses do not occur is found from $\alpha r\sin\vartheta_2=s_1+s_2$. The effect of the contact point at $\vartheta_1$ is still present, but the extra contact point at $\vartheta_2$ will introduce extra terms in the expression for the tail. We have
\[
\Upsilon(l) = \frac{4\!\sqrt{2s_1}}{5vs_2^2} (s_1+s_2-l)^{5/2} + \cdots , \quad l\longrightarrow (s_1+s_2)^-
,
\]
and $\Upsilon(l)=0$ for $l\geq s_1+s_2$. The term that needs to be added to (\ref{expand}) has the same form. It can be obtained from (\ref{expand}) by the replacements $\vartheta_1\longrightarrow\vartheta_2$ and $s_2\longrightarrow -s_2$. The power-law behavior of the Fourier tails for the different types of eclipses I, II, and III is given in Table \ref{table3}.
\end{document}